# Influence of Small Molecule Property on Antibody Response

Kai Wen[†], Yuchen Bai[†], Yujie Wei, Chenglong Li, Suxia Zhang, Jianzhong Shen &Zhanhui Wang[*]

*Beijing Advanced Innovation Center for Food Nutrition and Human Health, College of Veterinary Medicine, China Agricultural University, Beijing Key Laboratory of Detection Technology for Animal–Derived Food Safety, and Beijing Laboratory of Food Quality and Safety Beijing 100193, China*

Antibodies with high titer and affinity to small molecule are critical in the field for the development of vaccines against drugs of abuse, antidotes to toxins and immunoassays for compounds. However, little is known regarding how properties of small molecule *per se* influence and which chemical descriptor could indicate the degree of the antibody response. Based on our previous study, we designed and synthesized two groups of small molecules, called haptens, with varied hydrophobicities to investigate the relationship between properties of small molecules and antibody response in term of titer and affinity. We found that the magnitude of the antibody response is positively correlated with the degree of molecular hydrophobicity and related chemical descriptors. This study provides insight into the immunological characteristics of small molecules themselves and useful clues to produce high–quality antibodies against small molecules.

**Small molecule, property, antibody response, hydrophobicity, molecular descriptor, molecular polarity index**

Since the last century, antibodies have been highly successful in biomedicine, biochemistry and bioanalysis [1]. Antibodies are glycoproteins derived from the humoral immune naturally responding to macromolecules such as bacteria, viruses and proteins. In the case of low–molecular-weight compounds (<1000 Da), such as pharmaceuticals, poisons and drugs of abuse, these small molecules cannot be effectively recognized by the immune system to induce a subsequent antibody response [2]. To produce antibodies against small molecules, structural analogs of targets, named haptens, should be conjugated to larger molecules, called carriers. As well–established biological reagents, thousands of antibodies against hapten molecules with varied titers and affinities have been produced [3]. Unlike antibodies to macromolecules, desired antibodies to haptens should possess not only high titer but also high affinity for the free molecule, which is essential for their use as biopharmaceutics, antidotes and immunoprobes. Some factors, including hapten design, coupling chemistry, hapten-carrier ratio, carrier type and adjuvant formulation, have been reported to influence antibody response and then antibody formation, as they affect presentation of the hapten molecule and thus the quality of the immune response [4]. Nowadays, little is known regarding what primarily guarantees a high–quality antibody response to a hapten. In the author's opinion and experience, the decisive factor for antibody response should be the hapten itself; however, how hapten properties themselves influence and which molecular descriptors can indicate or even predict the degree of the antibody response is poorly understood.

To address this question, it is critical to deepen our understanding of the immunogenicity of small molecules, of course, after conjugating to carriers, which will be useful to produce the desired antibody with excellent titer and affinity. Systematic investigation of the relationship between properties of small molecules and antibody response is experimentally challenging due to the large number of variables and the limited number of typical molecular models. Recently, we successfully produced the antibodies against trimethoprim by preferentially exposing the hydrophilic dimethamine of trimethoprim, resulting in a surprisingly high titer and affinity (Figure 1 and Table S1) [5]. Compared with the antibodies obtained by exposing the hydrophobic trimethoxyphenyl of trimethoprim in the other reports, the affinity (expressed as $IC_{50}$ values) of our antibodies are improved by 10–90–fold with higher titers [6].

---
[†]These authors contributed equally to this work.

[*]Corresponding authors (email: wangzhanhui@cau.edu.cn)

These results reveal that the hydrophilic groups of haptens seem to account more for a strong antibody response than the hydrophobic groups of haptens. However, recent studies on the immunogenicity and haptenic character of poly (ethylene glycol) (PEG) and other nanoparticles showed that hydrophilic materials usually bear a lower immunogenicity than their hydrophobic counterparts [7]. These opposite results motived us to examine whether the capacity of small molecule to induce antibody response is related with its hydrophobic degree. If not, which of its molecular descriptors could suitably indicates the immunogenicity of small molecule? Since the small molecule itself has the most critical impact on antibody response in the author's opinion.

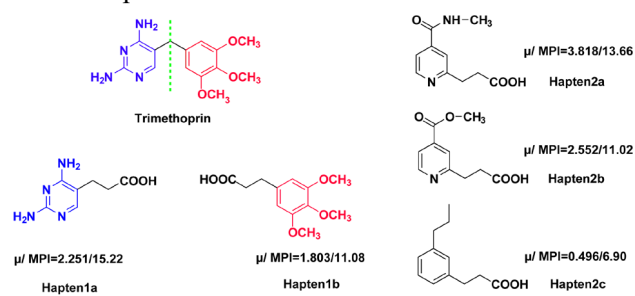

**Figure 1.** Chemical structures of haptens. Hapten1a mimics the trimethoxyphenyl of trimethoprim and is synthesized by us. Hapten1b mimics the diaminopyrimidine of trimethoprim and is a commercial compound. Hapten2 group were intentionally designed and synthesized with significantly different hydrophobicities. The molecular descriptors of haptens which excluding the spacer arm were calculated. Dipole moment, μ; molecular polarity index, MPI.

To understand the immunogenic attributes of small molecules, a simple and logical experiment based on simplified models is necessary. According to the immunizing results of trimethoxyphenyl, we firstly divided trimethoprim into two parts and separately verified the capacity of two resultant haptens to induce an antibody response. As shown in Figure 1, the proposed Hapten1a (in red) mimics the hydrophilic dimethamine moiety of TMP and Hapten1b (in blue) resembles the hydrophobic trimethoxyphenyl moiety of TMB. Secondly, we intentionally designed Hapten2 group with simple structure, which comprised three novel compounds with similar molecular weights and obviously different hydrophobicities (Figure 1). Based on the structural characteristics of three compounds, the order of hydrophobicity is expected to be: Hapten2c > Hapten2b > Hapten2a. The key point in the design of these haptens was the use of the same spacer arm, thus providing structural uniformity to remove the usual "bridge effects" arising from the linker between the hapten and carrier protein [8]. In addition, the similar molecular weight of members of the Hapten2 group ensured a fair comparison, since molecular weight is thought to be an important factor in immunogenicity of small molecule [9].

To ensure the rationality and practicability of these haptens designed, we firstly evaluate their hydrophobicity by using calculation chemistry with the M06–2X density functional and TZVP basis set before synthesis [10]. After obtaining the minimum conformations of the haptens excluding the spacer arm (Figure S4), the maps of electrostatic potential (ESP) mapped *van der Waals* surface of all haptens are constructed since ESP are closely related to the hydrophobicity of small molecules and could be as an indicator [11]. In addition, we calculated area distribution of different ESP intervals on the *van der Waals* surface. As shown in Figure 2A, the neutral regions in white of Hapten1a account for a large proportion of *van der Waals* surface due to the trimethoxyphenyl. In Hapten2，the hapten2c present predominantly neutral with some negative regions at the position of phenyl, is different from hapten2a and hapten2b which have obviously positive ESP in red at amino group and carboxyl group respectively (Figure 2B). More detail, the ESP surface area of Hapten1a is equably distributed between -30 kcal/mol and 30 kcal/mol while that of Hapten1b is concentrated in a narrower range of 0-20 kcal/mol which illustrates a more even surface distribution, roughly implying that Hapten1a possessing significantly lower hydrophobicity than Hapten1b. Similarly, in the group of Hapten2, by comparing the uniformity of ESP distribution, Hapten2c provides the highest molecular hydrophobicity followed by Hapten2b and then Hapten2a (Figure 2B). The calculated hydrophobicity of haptens support the practicability of these haptens we designed to study the influence of molecular hydrophobicity on antibody response. After confirmation by computational chemistry, the hapten1a and three haptens in Hapten2 group were totally synthesized in four or five steps (Scheme 1, 2, 3) since the hapten1b is commercially available. The detailed synthesis and characterization of haptens were provided in Supporting information (Figure S1).

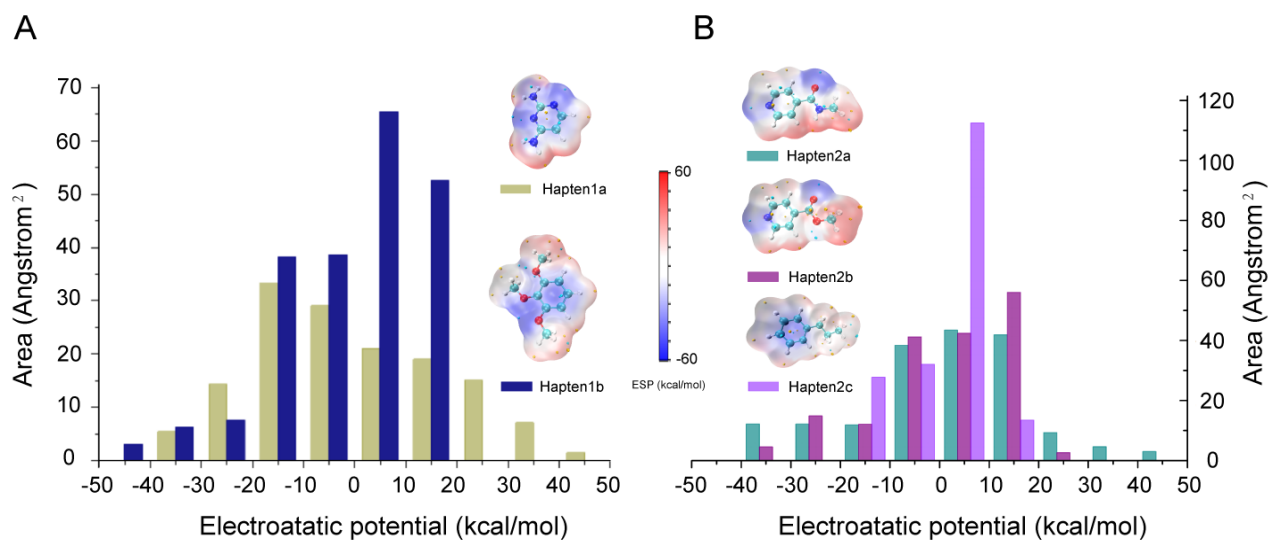

**Figure 2.** Electrostatic potential (ESP) mapped *van der Waals* surface (i.e. ρ = 0.001 a.u. isosurface) and area distribution of ESP intervals of (A) Hapten1 and (B) Hapten 2. The negative ESP regions are indicated in blue, the positive regions in red and the neutral regions in white.

To conjugate the haptens on a carrier protein, carboxylic acids of all hapten molecules were activated with *N*–hydroxysulfosuccinimide and dicyclohexylcarbodiimide and then treated with bovine serum albumin (BSA) as immunogens and ovalbumin (OVA) as coating antigens. Since the hapten–carrier ratio could influence the antibody response in animals and the subsequent antibody test [4b, 12], it was controlled carefully in the reasonable range of 13.4 to 17.0 for hapten–BSA and 4.4 to 5.3 for hapten–OVA, respectively (Figure S2–S3 and Table S2). Each hapten–BSA conjugate was used to immunize a set of 10 BALB/c mice with a total of three times at 25–day intervals. The use of 10 mice *per* hapten minimized the risk of obtaining results overly biased by any individual mouse (Table S4). Antisera are composed of the entire distribution of IgG antibodies present in mouse and provide a comprehensive view of general trends associated with the overall antibody response.

Thus, antisera titer and affinity were assessed by both noncompetitive ELISA and competitive ELISA against the corresponding homologous coating antigen, respectively (described in Supporting information). In the experiment, antibody titer is defined as the dilution of antisera that furnished an $OD_{max}$ between 1.5 and 2.0, while antibody affinity is expressed as the inhibition ratio calculated according to an equation (Supporting information).

**Scheme 1.** Synthesis of Hapten1a

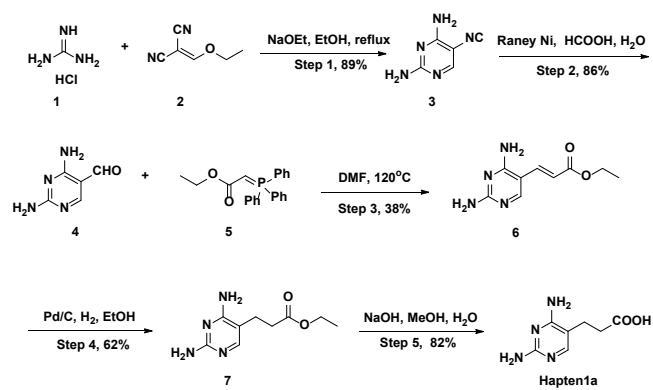

**Scheme 2.** Synthesis of Hapten2b and Hapten2a

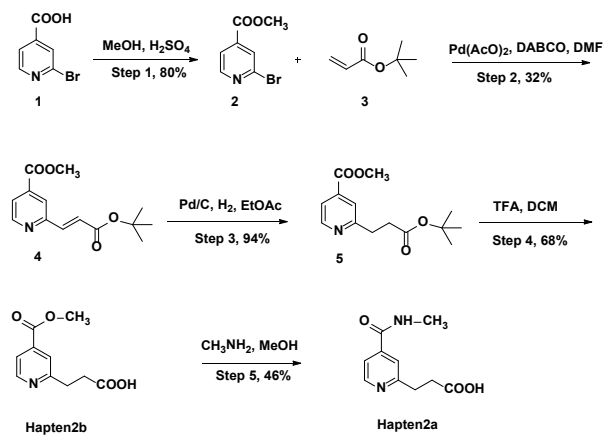

**Scheme 3.** Synthesis of Hapten2c

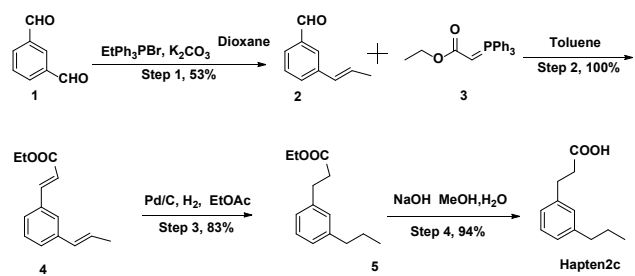

In short, a higher antibody dilution or inhibition ratio indicates a higher antibody response. As shown in Figure 3A, both Hapten1a and Hapten1b induced a significant antibody response. A general trend of gradually increasing titer as immunization continued was seen. Importantly, the antibody titers from Hapten1b were obviously higher than those from Hapten1a. During the immunization period, the antibody titers of Hapten1b were always 10–fold higher than those of Hapten1a, implying a stronger immune response achieved by more hydrophobic Hapten1b (Figure 3A). The performance of antibodies against small molecules should be judged by not only titer but also affinity. The latter is practically more important in the fields of bioanalysis, biochemistry and biomedicine in most cases. Unexpectedly, the antibodies induced by Hapten1a showed a rather poor affinity with an inhibition ratio of only 1%, unlike the antibodies induced by Hapten1b, which showed a greater than 75% inhibition ratio after the third immunization (Figure 3B). Furthermore, no improvement in the inhibition ratio of antisera for Hapten1a was observed, while the inhibition ratio of Hapten1b slowly increased with subsequent immunizations. These results reveal an obvious relationship between hydrophobicity of small molecule and antibody response, i.e., a higher degree of hydrophobicity of small molecule induced a stronger antibody response, which is opposite to our initial hypothesis supported by our previous study of trimethoprim [5].

To examine the unexpected relationship, three well–designed Hapten2 with varied hydrophobicity are then evaluated as shown in Figure 3C and 3D. The titers and affinities of all Hapten2 analogs increased sharply as immunizations continued, and an improvement in titer of at least 3–fold was achieved at the end of immunization. Similarly, a clear and definite relationship between hydrophobicity of small molecule and antibody titer was again observed with no exception. Moreover, both the highest antibody titer and the highest affinity of the antisera were obtained from the most hydrophobic hapten, Hapten2c. The titer and inhibition ratio of Hapten2c after the third immunization was approximately 7– and 10–fold, 2– and 3–fold higher those of Hapten2b and Hapten2a, respectively, demonstrating that the degree of small molecule hydrophobicity was positively correlated with the magnitude of the antibody response, at least in the case of the haptens we prepared. There is little difference in affinity between Hapten2a and Hapten2b, which may be related to similar ESP distribution on molecular surface of the two haptens (Figure 2B). These findings demonstrate the similarity between small molecules and polymers, nanomaterials and proteins, whose hydrophobicity significantly influences their antibody response [7]. Notably, the titers and affinities of all five evaluated haptens in this study were not high, and the $IC_{50}$ values for the haptens determined by ELISA were all at the μg/mL level (data not shown), which can be ascribed to the very low molecular weight and simple structures of these haptens.

To identify the appropriate molecular descriptors to indicate the ability of small molecule inducing antibody response, we extracted physicochemical descriptors of haptens such as LogP, molecular volume/surface, etc. (described in *Supporting information* and shown in Table S3) [10]. As shown in Table S3 and S5, only descriptors related to hydrophobicity, such as dipole moment, LogP, polar surface area and molecular polarity index (MPI), were observed to be directly proportional to the antibody response. MPI is defined for quantitatively measuring polarity of molecules based on ESP on their van der Waals surface, representing the uniformity of ESP distribution. Comparing to dipole moment, LogP, polar surface area, in the author's opinion, MPI maybe a better indicator of a hapten ability to induce an antibody response since it can be obtained easily and appropriately represents the molecular hydrophobicity in most cases [13].

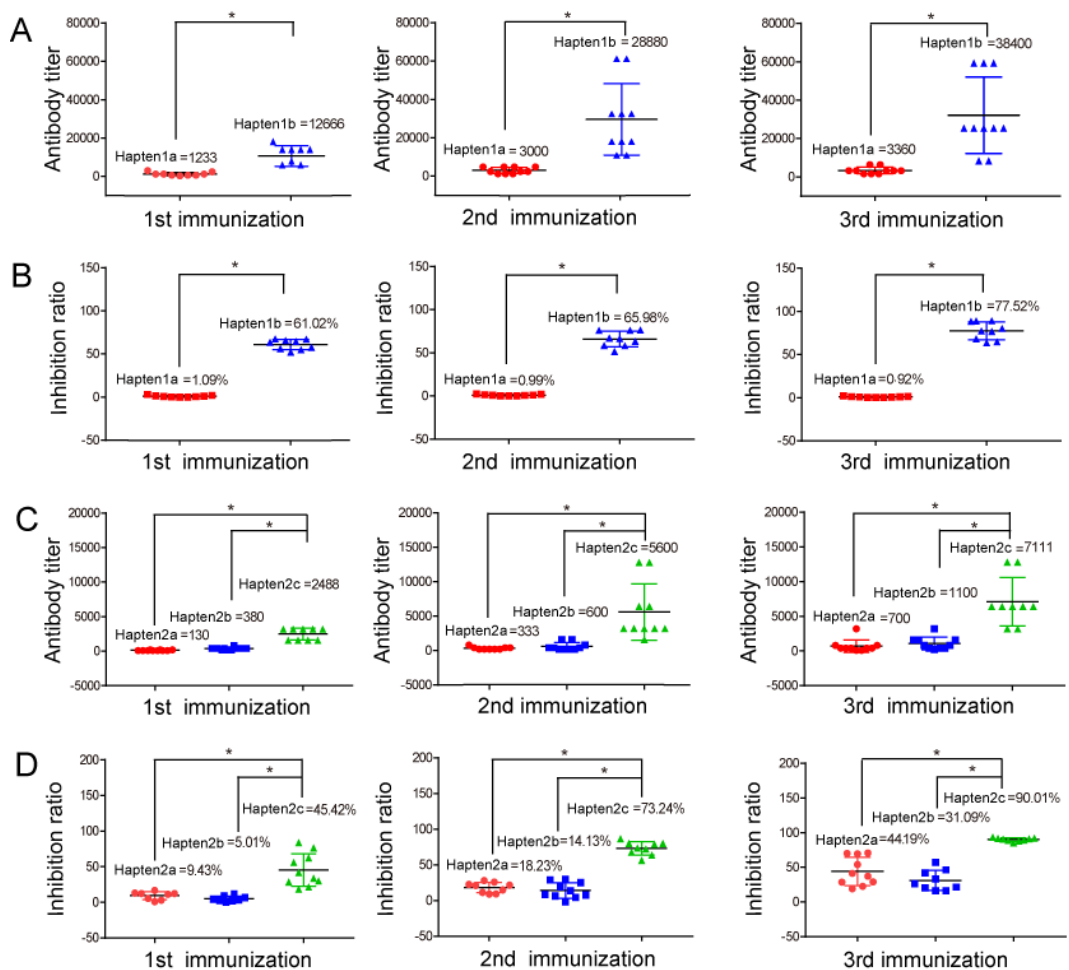

**Figure 3.** Evaluation of antibody titer and affinity. Antibody titers tested by homologous ELISA using serum from mice (n=10/group) at the 1st, 2nd and 3rd immunization. The titer is defined to be the dilution of antisera when furnishing ODmax between 1.5 and 2.0. Inhibition ratios representing antibody affinities measured by homologous ELISA for their corresponding hapten using serum from mice (n=10/group) at the 1st, 2nd and 3rd immunization. The inhibition ratio is calculated according to an equation described in supporting information. The higher inhibition ratio represents the higher antibody affinity. ＊P <0.05 determined by one–way ANOVA. (A) Antibody titer of Hapten1 (B) Inhibition ratio of Hapten1 (C) Antibody titer of Hapten2 (D) Inhibition ratio of Hapten2.

On the contrary, for example, the dipole moment cannot reflect the molecular hydrophobicities of symmetrical molecules, such as triacetone triperoxide and chlorhexidine. The correlation between MPI and antibody affinity was calculated by Person Correlation Matrix and the coefficient is -0.838 which is significant at the 0.05 level (2-tailed). The results clearly demonstrate that the MPI is a meaningful molecular descriptor to indicate the ability of small molecule inducing antibody response, at least in our study.

Despite this, a positive relationship between hapten hydrophobicity and antibody titer was not found between groups, for example, Hapten1a vs Hapten2a (Table S5). In addition, there have been many reports of exceptional small molecules with high hydrophobicity that produce a low antibody response, which here mainly refers to antibody affinity, such as squalene in comparison to its hydrophilic counterpart, etioporphyrin [14]. Furthermore, some small molecules with extremely high hydrophobicity did not induce an antibody response, such as diisononylcyclohexane–1,2–dicarboxylate and di–2–ethylhexyl terephthalate [15]. These results imply that an increase in hydro phobicity of small molecule does not always guarantee and improve antibody response, which is also influenced by other molecular properties of small molecules, such as size, shape, and flexibility, complicating the structure–magnitude correlation. Further study should be conducted by using more typical molecule models with strict control to obtain more precise relationship between molecular properties and antibody response.

In summary, we designed two groups of haptens with different hydrophobicities and systematically investigated the relationship between properties of small molecules and antibody response. A highly positive correlation between molecular hydrophobicity and antibody response was observed. The MPI was suggested to be used as an indicator of ability of small molecule inducing antibody response. These findings provide a deeper understanding of the antibody response of small molecule *per se* and have far–reaching implications for rational hapten design, a critical step toward the development of a new generation of vaccines against drugs of abuse, antidotes to chemical toxins and diagnostic antibodies for small molecules.

**Acknowledgements** This work was supported by the National Science Foundation of China (Grant No. 31873025).

**Conflict of interest** The authors declare that they have no conflict of interest.

**Supporting information** The supporting information is available online at…

# SUPPORTING INFORMATION

**Influence of Small Molecule Property on Antibody Response**

Kai Wen[†], Yuchen Bai[†], Yujie Wei, Chenglong Li, Suxia Zhang, Jianzhong Shen &

Zhanhui Wang[*]

*Beijing Advanced Innovation Center for Food Nutrition and Human Health, College of Veterinary*

*Medicine, China Agricultural University, Beijing Key Laboratory of Detection Technology for*

*Animal–Derived Food Safety, and Beijing Laboratory of Food Quality and*

*Safety Beijing 100195, China.*

# Table of Contents for Supporting Information

# Supplementary Schemes, Tables and Figures

**Table S1.** Previous reported haptens of trimethoprim and related conjugation chemistry, $IC_{50}$ values and titers of antibodies.

| Hapten structures | Conjugation chemistry | Antibody types | $IC_{50}$ (ng/mL) | Titers [a] | References |
|---|---|---|---|---|---|
| The following haptens reported in literatures which all expose diaminopyrimidine to immune system ||||||
| 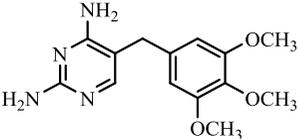 | Glutaraldehyde used as conjugation reagents | Polyclonal antibody from rabbit | >6.0[b] | 1,500 | 1 |
| | | | >6.0[b] | -[c] | 2 |
| | | Monoclonal antibody from mouse | 4.14 | 32,000 | 3 |
| | | | 1.98 | - | 4 |
| The following three haptens we prepared which much expose diaminopyrimidine moiety to immune system ||||||
| 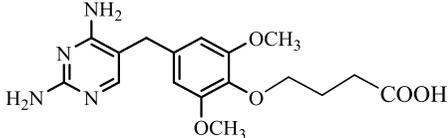 | Dicyclohexylcarbodiimide and $N$-hydroxysuccinimide used as conjugation reagents | Monoclonal antibody from mouse | 0.232 | 16,000 | 5 |
| | | | 0.067 | 60,000 | |
| | | | 0.2 | 20,000 | |
| 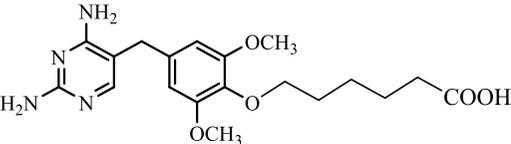 | Dicyclohexylcarbodiimide and $N$-hydroxysuccinimide used as conjugation reagents | | 0.15 | 40,000 | 6 |
| | | | 0.24 | 40,000 | |
| 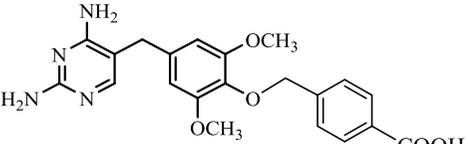 | Dicyclohexylcarbodiimide and $N$-hydroxysuccinimide used as conjugation reagents | | 0.17 | 60,000 | |

[a] The titer here was defined that the concentration of antibody generated the OD values ranged 1.5 to 2.0.

[b] The papers have not reported $IC_{50}$ values and only provided detection limits. Usually the $IC_{50}$ value was about 10 times of detection limit in an immunoassay.

[c] Not exhibited in the work.

1. **Materials and apparatus**

   Bovine serum albumin (BSA), horseradish peroxidase (HRP), *N*-hydroxysuccinimide (NHS), and 1-(3-dimethylaminopropyl)-3-ethylcarbodiimide (EDC) were purchased from Sigma-Aldrich (St. Louis, MO, USA). The substrate 3,3′,5,5′-tetramethylbenzidine (TMB) system was obtained from the Beijing WDWK Biotechnology Co., Ltd. (Beijing, China). All other chemicals and solvents were of analytical grade or better and were obtained from the Beijing Chemical Reagent Co. (Beijing, China). Reverse osmosis water, pyrogen free (RO water) was prepared using a Milli-Q water purification system (Millipore, Bedford, MA, USA).

2. **Buffers**

   The common buffer solutions used in the experiment are listed: (1) coating buffer (0.05 mol/L) carbonate buffer, pH 9.6); (2) phosphate buffer solution (PBS buffer, 0.01 mol/L, pH 7.4); (2) blocking buffer (2% skim milk powder (w/v)); (3) washing buffer (PBST, 10 mmol $L^{-1}$ PBS buffer with 0.05% Tween-20 (v/v)), pH 7.2); (4) antibody diluent buffer (10 mmol/L washing buffer with 0.2% albumin, (w/v)); (5) Goat anti-mouse IgG (HRP labeled) dilution buffer (PBS buffer containing 5% albumin (w/v)); (7) stopping reagent (2 mol/L $H_2SO_4$).

3. **Apparatus**

   Polystyrene microplates were obtained from Corning Inc (Costar 2592, Corning, NY, USA). Optical density (OD) was recorded on SpectraMax M5 (Downingtown, PA, USA). The chemical structures of all haptens were confirmed by liquid chromatography triple quadrupole mass spectrometry (Thermo, USA) and nuclear magnetic resonance spectrometry (NMR) DRX (Bruker, Rheinstetten, Germany). The conjugation ratio was determined by matrix-assisted laser desorption ionization time of flight mass spectrometry (MALDI–TOF-MS) (Bruker, Rheinstetten, Germany).

## 4. Synthesis and characterization of haptens

*4.1 Synthesis of Hapten1a*

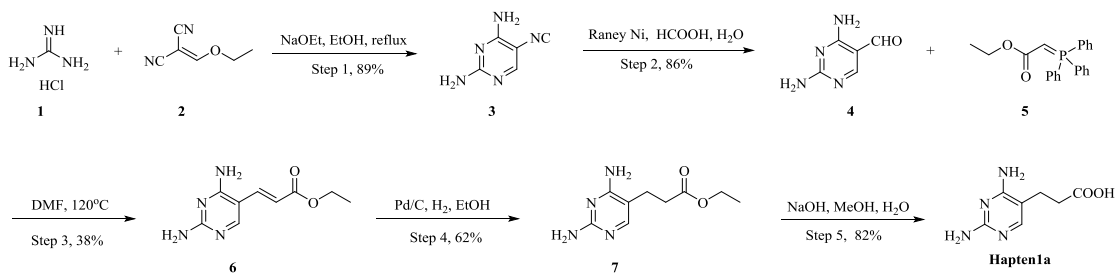

**Scheme 1**. The synthetic route of Hapten1a.

Hapten1a was prepared based on the following scheme in 5 steps.

*Step 1* Synthesis of compound 3 (2,4-diaminopyrimidine-5-carbonitrile)

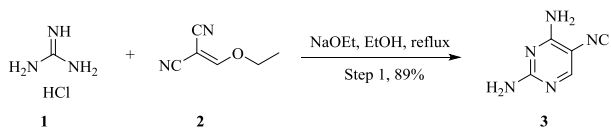

Guanidine hydrochloride (19.11 g, 200 mmol) was suspended in anhydrous ethanol (600 mL), 2-(ethoxymethylene)malononitrile (27.20 g, 200 mmol) was added in one portion. The mixture was stirred at ambient for 10 min. Sodium ethoxide (6.80 g, 100 mmol) was added in 2 portions in 10 min. The reaction mixture was heated at 81 °C (oil bath) for 2 h and then incubated at ambient temperature (25 °C) for 16 h. Solvent was removed in vacuum to give a yellow residue. Sat. ammonium chloride (500 mL) and pure water (100 mL) were added to the residue. The resulting mixture was stirred for 10 min, filtered, and then the desired product (24.01 g, 177.85 mmol, 89%). $^1$H NMR (300 MHz, DMSO-$d_6$) δ 6.93 (s, br, 2H) 7.10 (s, br, 2H), 8.17 (s, 1H) was obtained by drying the yellow solid in vacuum.

*Step 2* Synthesis of compound 4 (2,4-diaminopyrimidine-5-carbaldehyde)

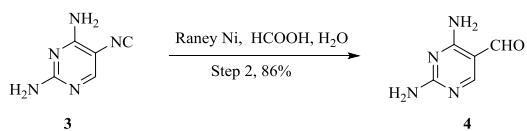

Compound 3 (24.01 g, 177.80 mmol) was added to a slurry of Raney nickel (50% in water, 20.00 g) followed by addition of formic acid (88% in water, 200 mL). The resulting mixture was then heated under reflux (125 °C, oil bath) for 7 h. Then stirred at ambient temperature for 16 h. The reaction mixture was filtered, the cake was washed with formic acid (100 mL). The filtrate was combined and concentrated to afford a yellow residue. The residue was poured onto ammonia (28%, 200 mL). The resulting mixture was stirred for 30 min. The yellow solid was collected by filtration, dried in hot air to gain a yellow solid (21.09 g, 86%) as desired product ($^1$H NMR (300 MHz, DMSO-$d_6$) δ 7.13 (s, 2H) 7.54 (s, br, 1H), 7.82 (s, br, 1H), 8.32 (s, 1H), 9.46 (s, 1H)).

***Step 3*** Synthesis of compound 6 （(E)-ethyl 3-(2,4-diaminopyrimidin-5-yl)acrylate）

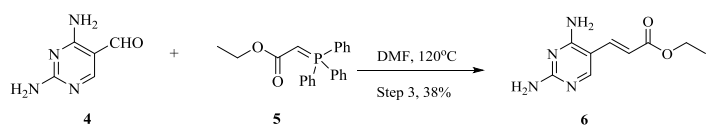

A suspension of compound **4** (5.52 g, 40 mmol), compound **5** (16.70 g, 48 mmol) in DMF (100 mL) was heated at 120 °C for 4 h. The mixture was cooled to room temperature. Water (400 mL) was added to the residue then stirred for 30 min. Yellow precipitate appeared. The mixture was extracted with ethyl acetate (400 mL × 4). The combined organic layers were washed with water (600 mL), brine (600 mL), dried Na$_2$SO$_4$. Solvent removal under vacuum gave a yellow powder (19.02 g). The powder was washed with a mixture of petroleum ether and ethyl acetate (1:1, 200 mL × 3) and concentrated to give a powder as desired product (3.20 g, 38%).

***Step 4*** Synthesis of compound 7 （ethyl 3-(2,4-diaminopyrimidin-5-yl)propanoate）

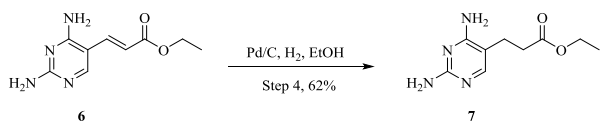

A mixture of compound **6** (3.20 g, 15.30 mmol), palladium on carbon (10%, 1.00 g) in ethanol (150 mL) was hydrogenated at 50 °C under a hydrogen balloon for 48 h. The mixture was cooled to room temperature. Palladium on carbon was removed by filtration. The filtrate was concentrated to afford a residue (3.10 g). The mixture was refluxed in a mixture of MeOH and CHCl$_3$ (1:3, 100 mL) and filtered. The solid was discarded. The filtrate was concentrated to give a residue, which was washed with a mixture of petroleum ether and dichloromethane (10:1, 40 mL × 2) to give a residue as desired crude product (2.10 g, 62%). The material was used in next step without further purification.

*Step 5* Synthesis of Hapten1a

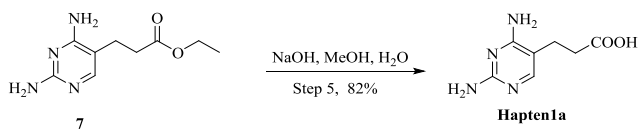

To a suspension of compound 7 (2.10 g, crude, 10 mmol) in MeOH (15 mL) was added a solution of NaOH (2.40 g, 60 mmol) in water (15 mL). The resulting mixture was stirred at room temperature for 2 h. TLC monitoring indicated that compound 7 was consumed completely. The mixture was neutralized with HCl (3 M, 20 mL) to adjust pH value to 3-4. The mixture was extracted with dichloromethane (20 mL × 3). The aqueous layer was concentrated to give a residue. The residue was refluxed in a mixture of MeOH and water (5:1, 60 mL) and filtered while hot. The filtrate was concentrated to give a residue. The residue was refluxed in a mixture of MeOH and water (10:1, 55 mL) and filtered while hot. The filtrate was concentrated to give an off-white solid as pure target compound (1.50 g, 82%) (1H NMR (300 MHz, D$_2$O) δ 2.65-2.67 (m, 4H), 7.50 (s, 1H). ESI-MS found 181 (M-H)).

*4.2 Synthesis of Hapten2a and Haten2b*

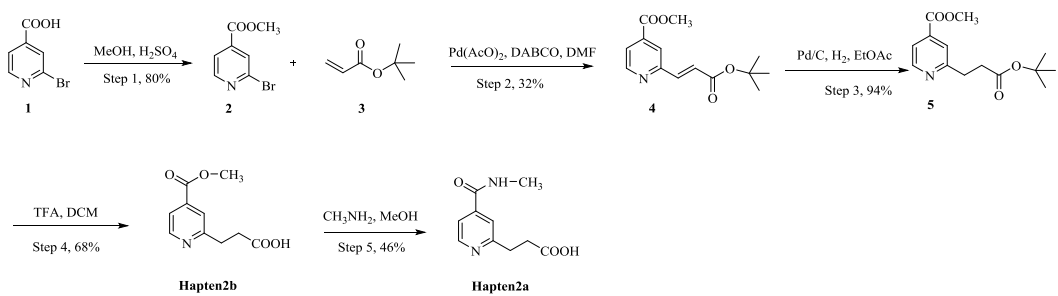

**Scheme 2**. The synthetic route of Hapten2a and Hapten2b.

Target Hapten2a and Hapten 2b was prepared based on the following scheme.

*Step 1* Synthesis of compound 2 (methyl 2-bromoisonicotinate)

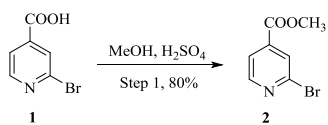

Concentrated sulphate acid (2 mL) dropwise was added to a mixture of compound **1** (20.20 g, 100 mmol) in MeOH (200 mL). The mixture was heated under reflux for 5 h. The mixture was cooled to room temperature. Ethyl acetate (300 mL) was then added. The mixture was washed with sat. NaHCO$_3$ to neutral. The mixture was then dried over sodium sulphate. The solvent was removed under vacuum to give a light-yellow oil as desired compound 2 (17.20 g, 80%).

*Step 2* Synthesis of compound 4 ((E)-methyl 2-(3-tert-butoxy-3-oxoprop-1-enyl)isonicotinate)

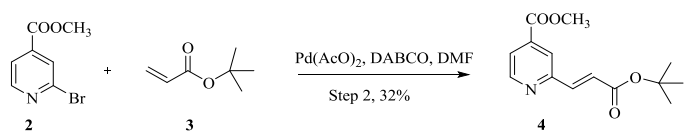

Under nitrogen atmosphere, to a mixture of compound **2** (2.16 g, 10 mmol), potassium carbonate (1.65 g, 12 mmol), triethylenediamine (112.30 mg, 1 mmol), Bu$_4$NBr (3.87 g, 12 mmol) was added DMF (20 mL) via syringe. Then Pd(OAc)$_2$ (112.3 mg, 0.50 mmol) and compound **3** (3.20 g, 25 mmol) were added. The mixture was heated at 120 $^o$C under nitrogen atmosphere for 6 h. The mixture was cooled to room temperature. Water (100 mL) was then added. The mixture was extracted with dichloromethane (50 mL × 3). The combined organic layers were dried over Na$_2$SO$_4$. The mixture was then concentrated. The residue obtained was subject to silica gel column separation (petroleum ether : ethyl acetate=30 :1) to give a light-yellow solid (880 mg, 32 %) as desired compound **4.**

*Step 3* Synthesis of compound 5 (methyl 2-(3-tert-butoxy-3-oxopropyl)isonicotinate)

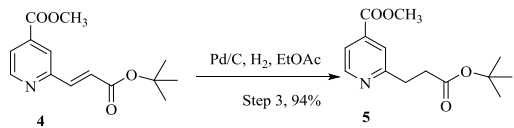

To a solution of compound 4 (5.00 g, 19 mmol) in ethyl acetate (60 mL) was added palladium on carbon (10%, 2.00 g). The mixture was hydrogenated under hydrogen gas balloon for 16 h. The mixture was then filtered to remove the palladium on carbon. The filtrate was concentrated to give a light-yellow oil as desired compound 5 (4.73 g, 94%).

*Step 4* Synthesis of Hapten2b

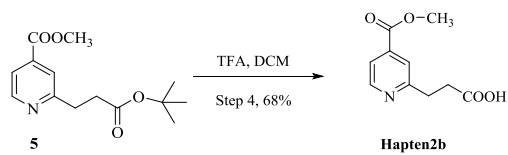

TFA (10 mL) was added to a solution of compound **5** (4.73 g, 17.80 mmol) in dichloromethane (30 mL). The mixture was then refluxed for 4 hours. The mixture was cooled to room temperature. Solvent was removed under vacuum. To the mixture was added water (50 mL) and the pH value was adjusted to 4 with the addition of 1 N HCl. The precipitate was filtered and collected and dried under vacuum to give desired Hapten2b as a white solid (2.53 g, 68%). $^1$H NMR (400 M Hz, DMSO-d6) δ2.70 (t, J = 7.2 Hz, 2H), 3.07 (t, J = 7.2 Hz, 2H), 3.90 (s, 3H), 7.65-7.74 (m, 2H), 8.69 (d, J = 4.8 Hz, 1H), 12.13 (s, 1H).

*Step 5* Synthesis of Hapten2a

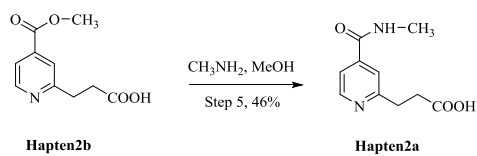

To a mixture of Hapten2b (3.78 g, 18 mmol) in MeOH (40 mL) cooled to 0 °C was added a solution of methyl amine (4.41 g, 30%, 90 mmol). The mixture was then heated at 50 °C for 8 h. The mixture was cooled to room temperature and concentrated. Water (50 mL) was added. The pH value of the mixture was adjusted to 4 with the addition of 1 N HCl. The precipitate was collected via filtration to give a solid. Further drying under vacuum gave a white solid as desired product **3** (1.72 g, 46%). $^1$H NMR (500 MHz, DMSO-$d_6$) δ 2.71 (t, $J$ = 7.0 Hz, 2H), 2.80 (d, J = 4.5 Hz, 3H), 3.04 (t, $J$ = 7.0 Hz, 2H), 7.56 (dd, $J$ = 1.5 Hz, 5.5 Hz, 1H), 7.64 (s, 1H), 8.60 (d, $J$ = 5.0 Hz, 1H), 8.69 (d, $J$ = 4.5 Hz, br, 1H), 12.08 (s, br, 1H).

## 4.3 Synthesis of Hapten2c

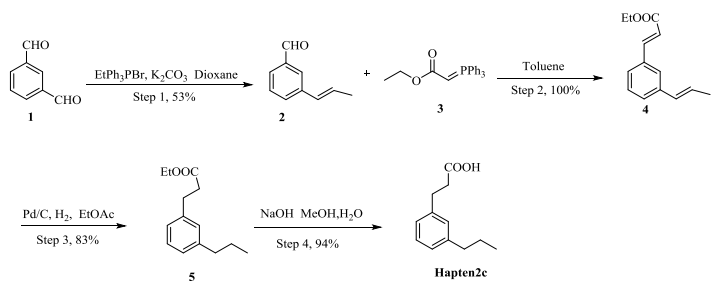

**Scheme 3**. The synthetic route of Hapten2c

Target Hapten3c was prepared based on the following scheme in 4 steps

***Step 1*** Synthesis of compound 2 ((E)-3-(prop-1-enyl)benzaldehyde)

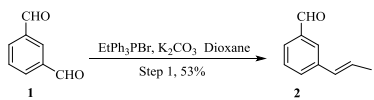

A mixture of compound 1 (2.68 g, 20 mmol), ethyltriphenylphosphonium bromide (8.20 g, 22 mmol), potassium carbonate (4.20 g, 30 mmol) in dioxane (30 mL) and water (1 mL) was heated under reflux for 8 h. The reaction mixture was cooled to room temperature. Solvent was removed under vacuum. To the residue was added water (50 mL) and extracted with ethyl acetate c (50 mL × 3). The combined organic layers were dried over sodium sulphate. Solvent removal gave a residue. The residue was subject to silica gel column chromatography (petroleum ether: ethyl acetate = 50 :1) to give a yellow oil (1.54 g, 53 %) as desired compound **2.**

***Step 2*** Synthesis of compound 4 ((E)-ethyl 3-(3-((E)-prop-1-enyl)phenyl)acrylate)

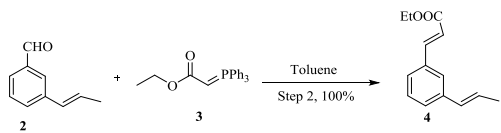

A mixture of compound 2 (730 mg, 5 mmol) and compound 3 (1.93 g, 5.75 mmol) in toluene was refluxed for 3 h. The mixture was cooled down and concentrated. The residue was purified with silica gel column (petroleum ether: ethyl acetate = 10 :1) to give a colorless oil as desired compound 4 (1.17 g, 100%).

***Step 3*** Synthesis of compound 5 (ethyl 3-(3-propylphenyl)propanoate)

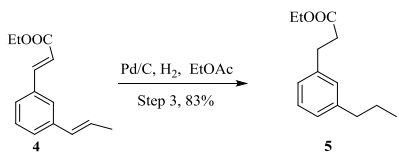

To a solution of compound 4 (1.10 g, 5.44 mmol) in ethyl acetate (15 mL) was added palladium on carbon (10%, 200 mg). The reaction mixture was hydrogenated under hydrogen gas balloon for 16 h at room temperature (25 °C). The mixture was filtered to remove the palladium on carbon. The filtrate was concentrated to give desired products compound 5 as a yellow oil (1.02 g, 83%).

***Step 4*** Synthesis of Hapten2c

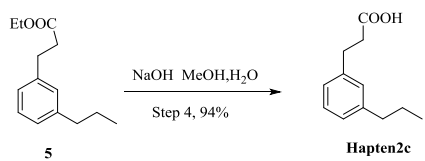

To a suspension of compound 5 (3.00 g, 13.60 mmol) in MeOH (30 mL) was added water (5 mL) and NaOH (1.97 g, 49 mmol). The mixture was then heated at 80 °C for 2 h. The mixture was then cooled to room temperature. MeOH was removed under vacuum. To the reaction mixture was added water (20 mL). The pH value of the mixture was adjusted to 2 via addition of 1 N HCl. The mixture was extracted with ethyl acetate (50 mL × 3). The combined organic layers were dried over sodium sulphate. Solvent removal under vacuum gave a light yellow oil as desired target (2.48 g, 94%).[1]H NMR (400 MHz, CDCl$_3$) δ 0.93 (t, *J* = 5.7 Hz, 3H), 1.58-1.67 (m, 2H), 2.55 (t, *J* = 5.7 Hz, 2H), 2.67 (t, *J* = 5.7 Hz, 2H), 2.93 (t, *J* = 6.0 Hz, 2H), 7.01-7.03 (m, 3H), 7.19-7.21 (m, 1H), 11.32 (s, br, 1H).

(A)

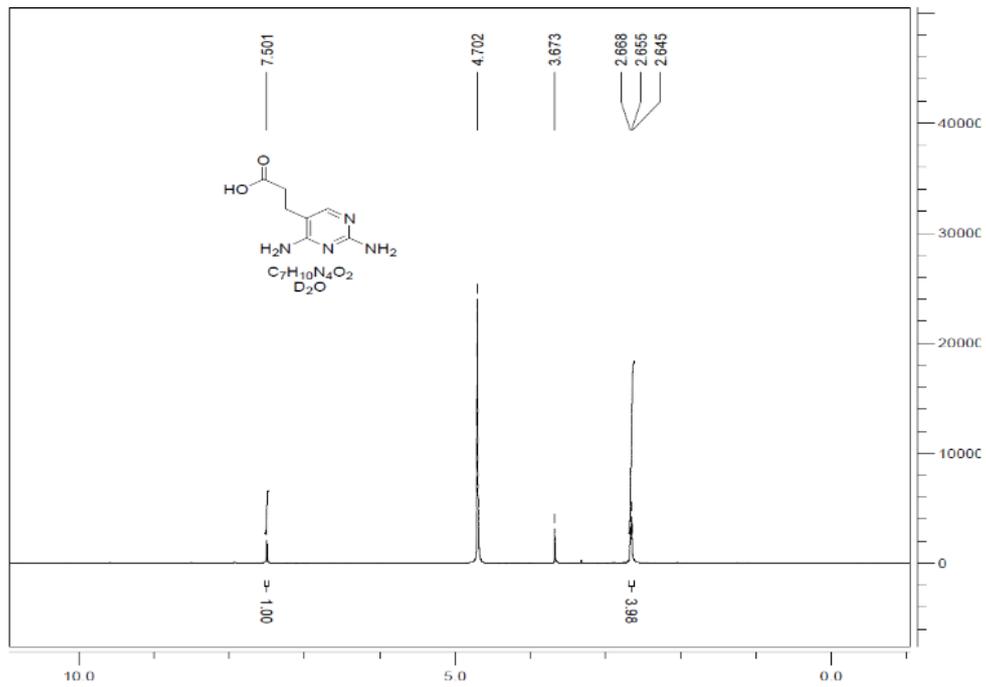

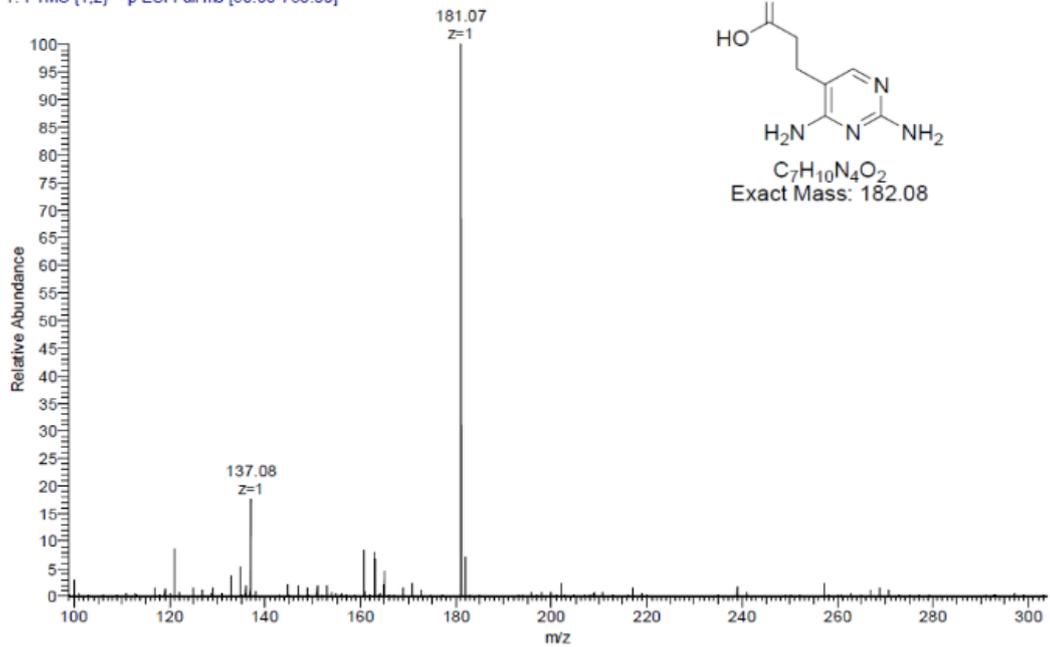

**( B )**

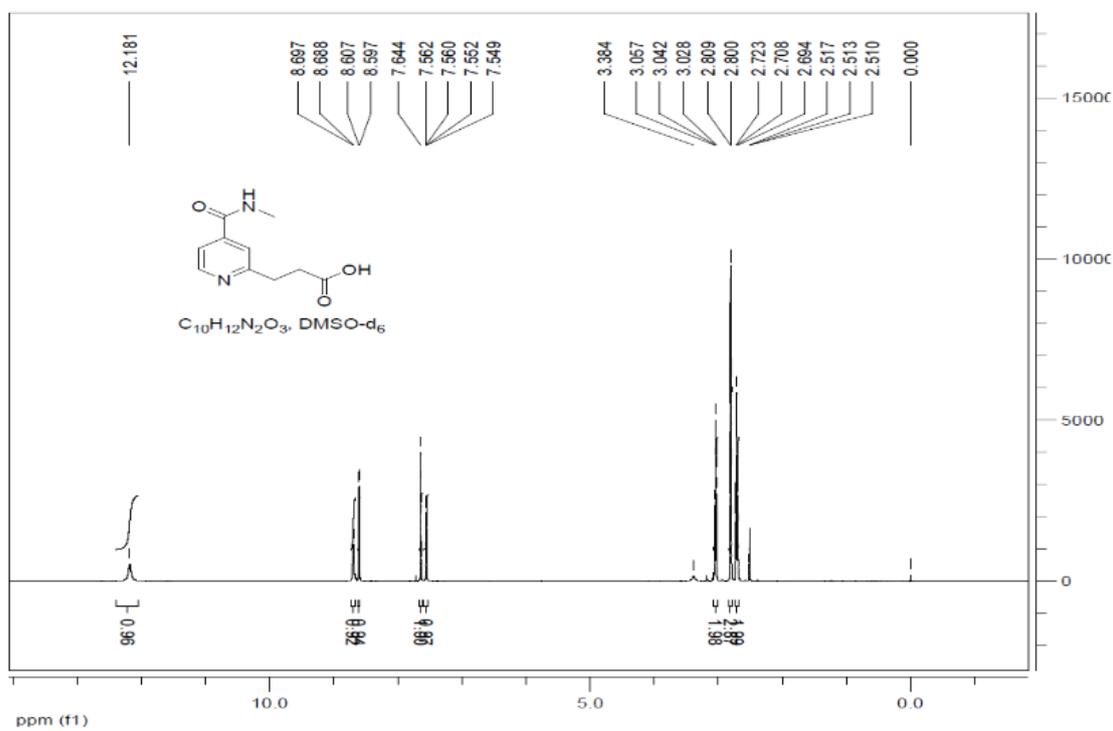

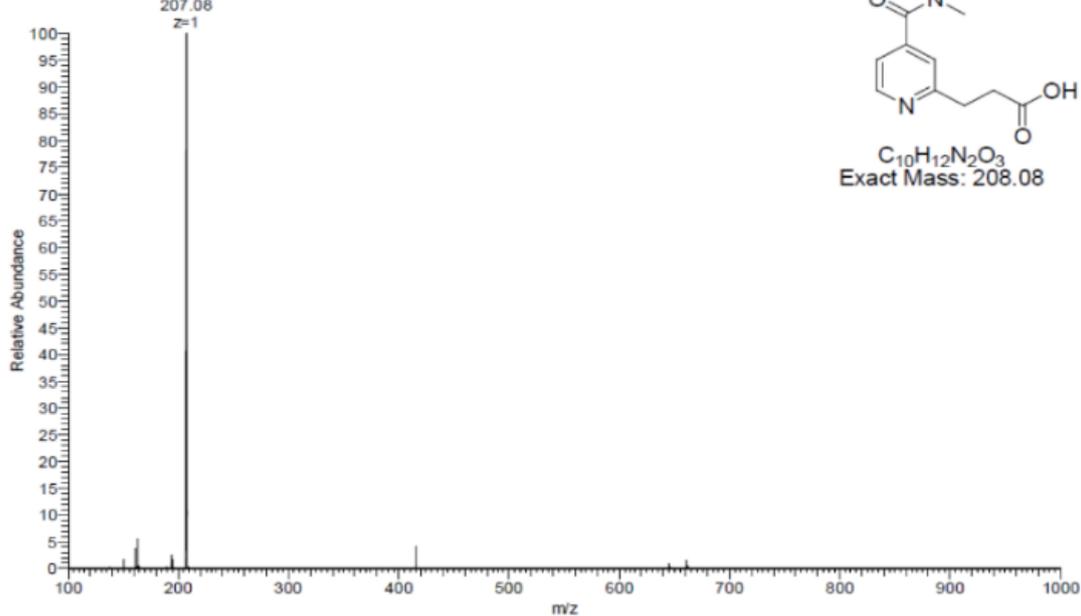

**( C )**

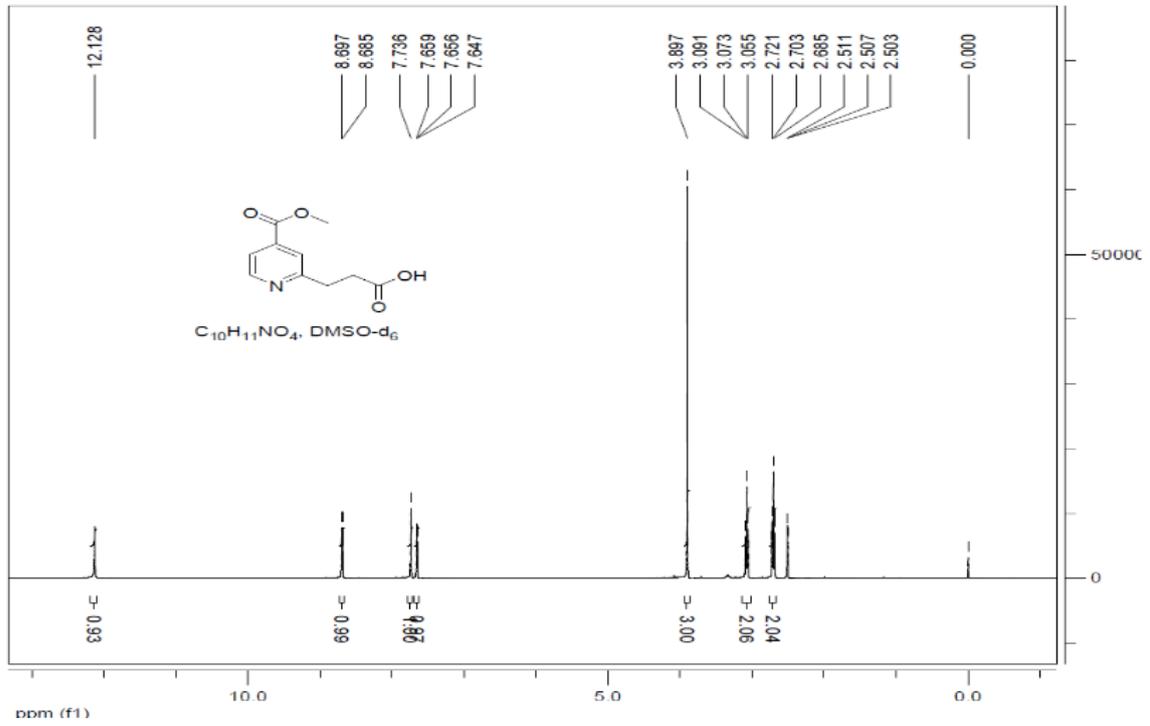

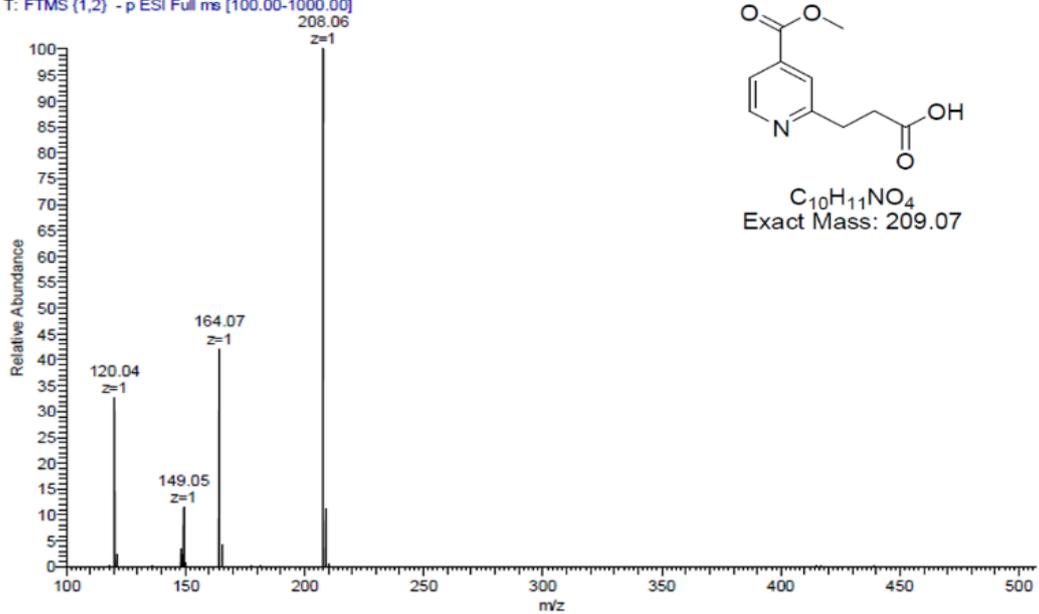

( D )

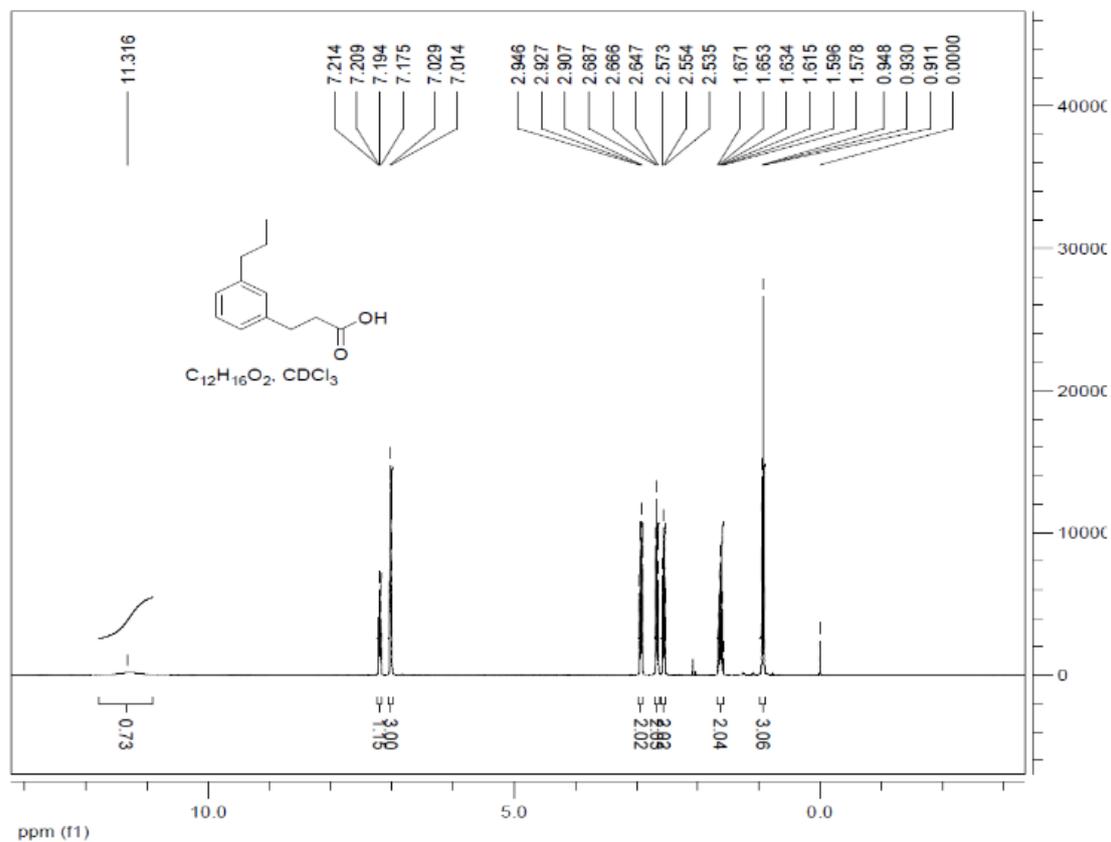

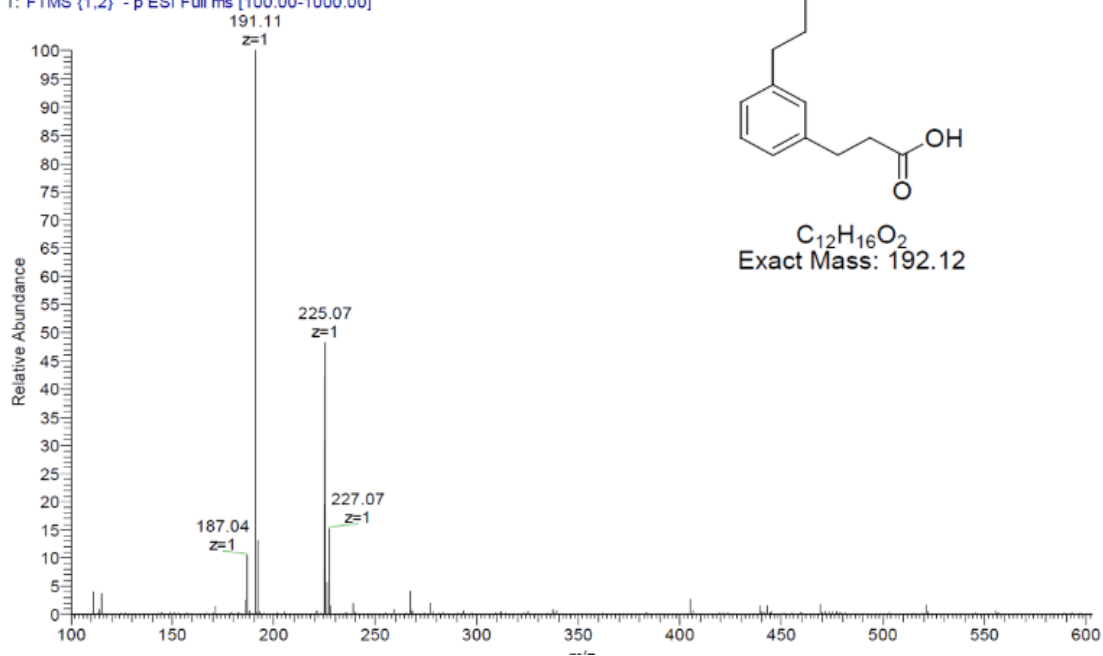

**Figure S1**. The mass spectra and ¹H NMR spectra of (A) Hapten1a, (B) Hapten2a, (C) Hapten2b, (D) Hapten2c.

## 5. Preparation and characterization of conjugates

The haptens with carboxylic acid groups were activated through the active-ester method and then coupled to a protein (BSA or OVA). To control the hapten-protein ratios, different combination of haptens and protein were optimized (Table S1). For Hapten1a, 12.5 μmol, 14.6 μmol, 16.7 μmol, 18.7 μmol and 20.8 μmol of hapten1a was dissolved in 1.5 mL DMF, respectively, and NHS and DCC were added and mixed in 1.5 molar ratios. The mixture was reacted using a magnetic stirrer at room temperature for 6 h followed by the removal of the precipitates. The supernatant of mixture was dropwise added into dissolved BSA or OVA in 10 mL PBS in 1:75, 1:87.5, 1:100, 1:112.5, 1:125 ratio, respectively. In the case of other four haptens, 5.0 μmol, 5.8 μmol, 6.6 μmol, 7.5 μmol and 8.3 μmol of haptens were used and the molar ratio of DCC and NHS was 1:1.5. The molar ratios for hapten/protein were 1:30, :35, 1:40, 1:45, 1:50 respectively. All hapten-protein conjugates were stirred for 12 h and then dialyzed in PBS for 2 days. Hapten-BSA served as an immunogen and hapten-OVA acted as coating antigens were stored at −20 °C for further use. The conjugation ratios were characterized using MALDI–TOF-MS and were calculated as follows:

$$\text{Conjugation ratio} = \{M(\text{conjugates}) － M(\text{BSA})\}/M(\text{haptens}) \quad (Eq.1).$$

(A)

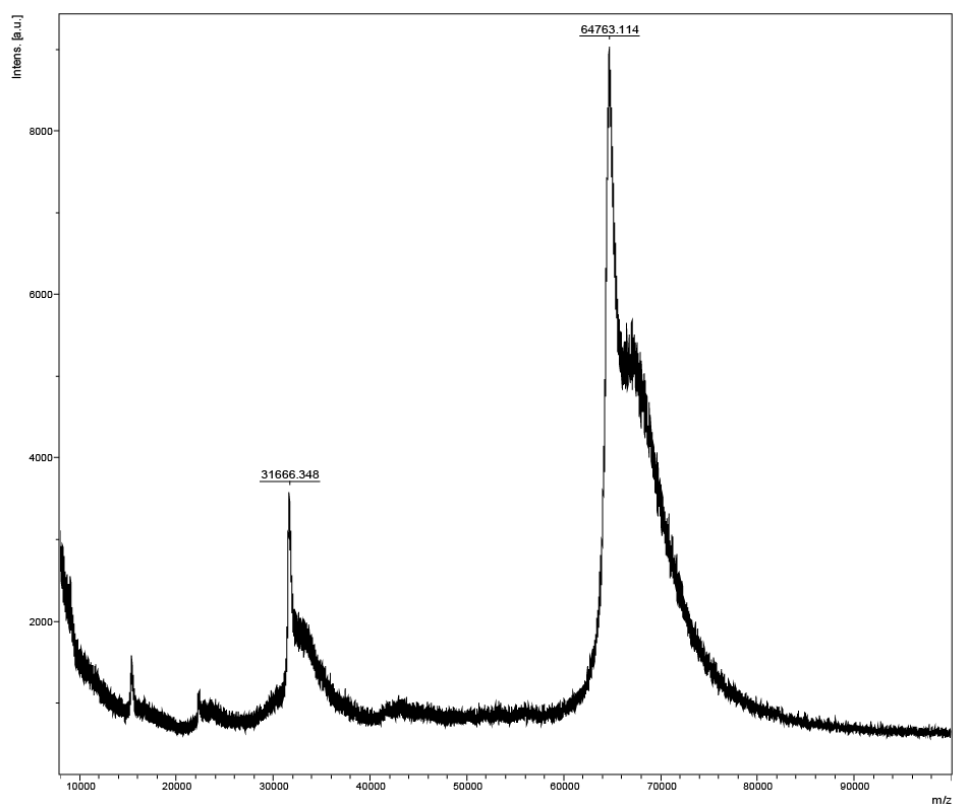

(B)

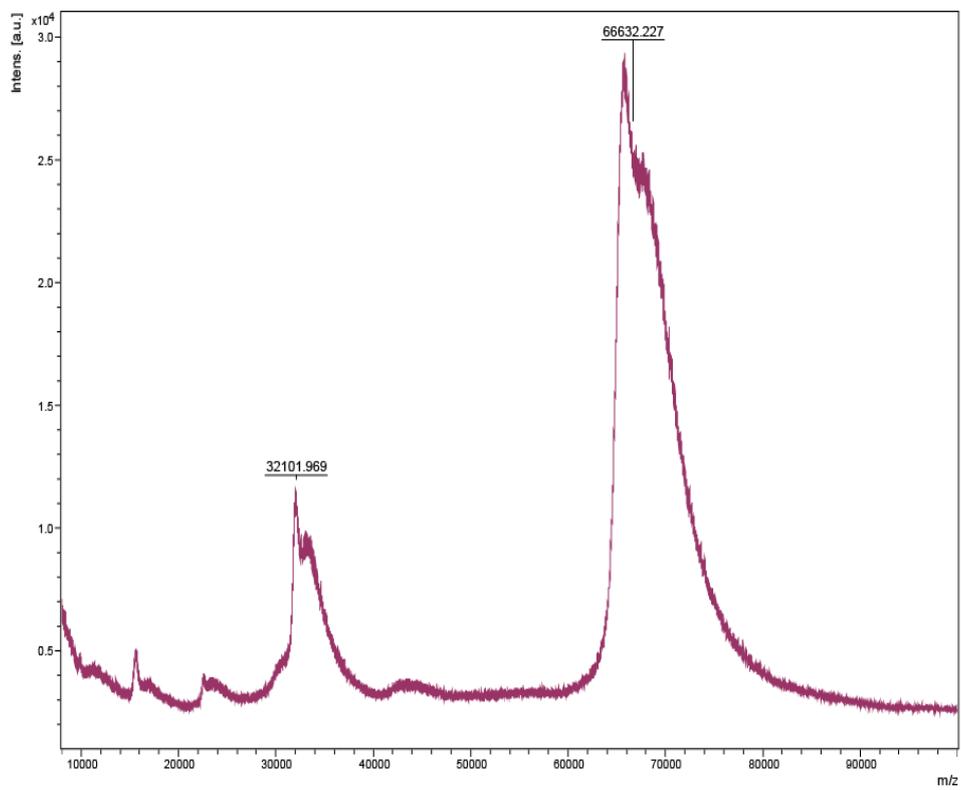

(C)

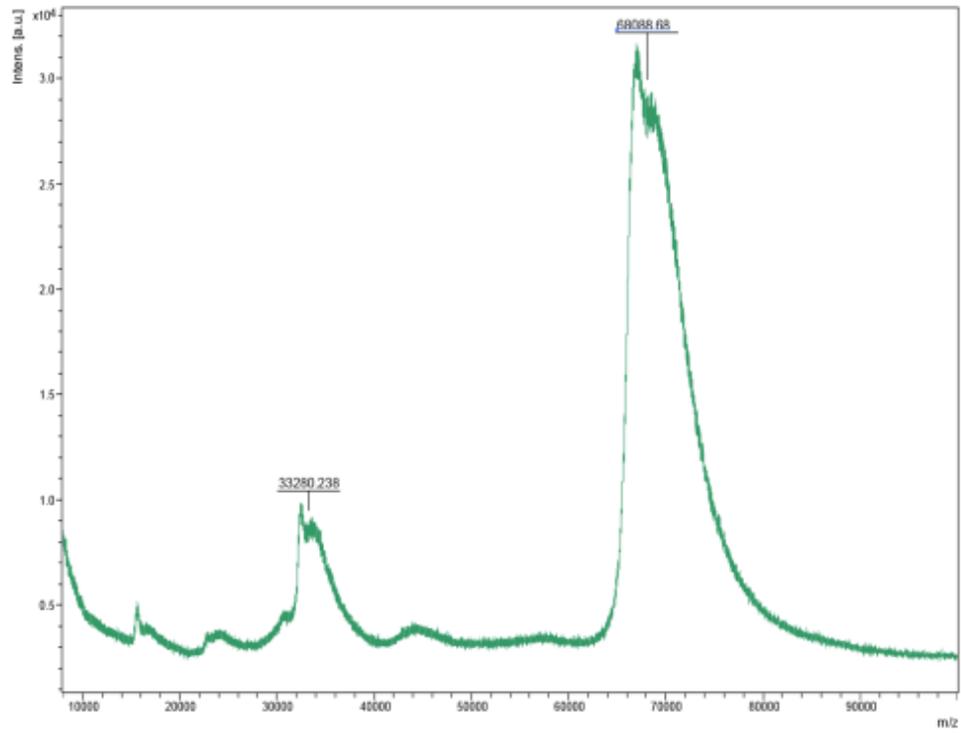

(D)

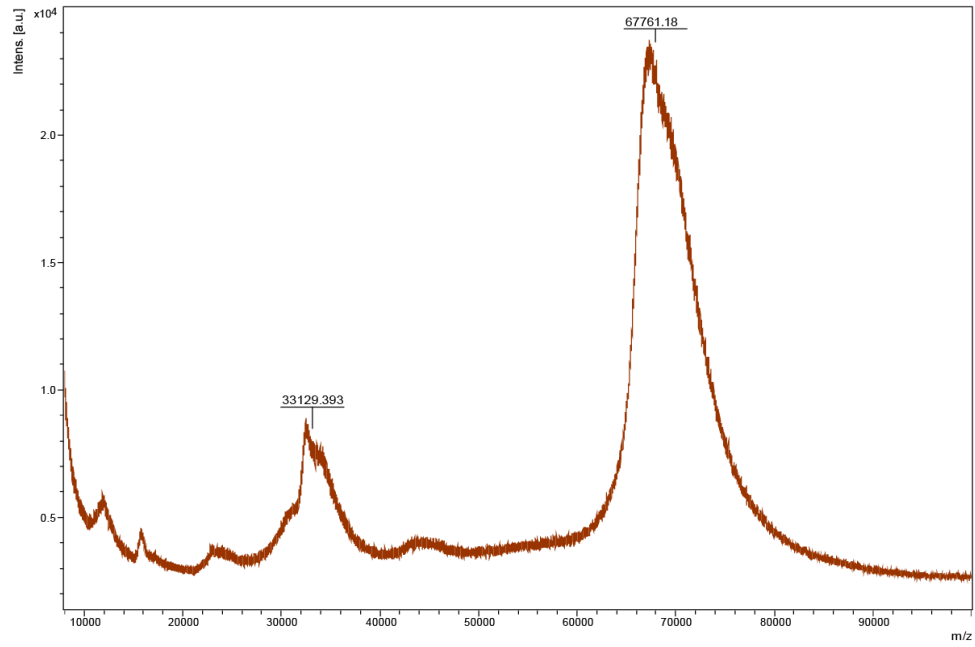

(E)

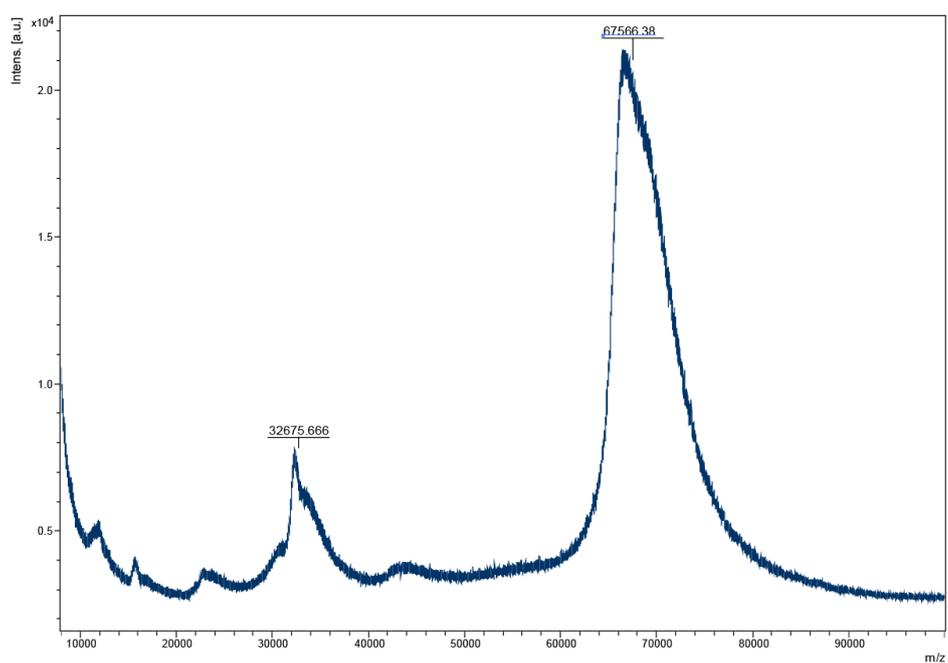

(F)

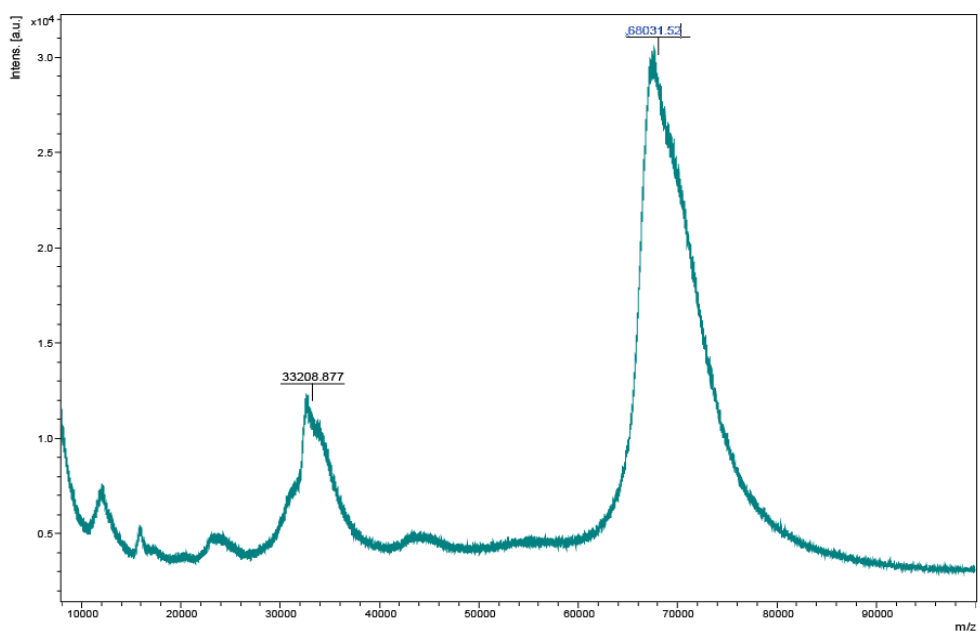

**Figure S2**. MALDI–TOF-MS of (A) BSA, (B) Hapten1a-BSA, (C) Hapten1b-BSA, (D) Hapten2a-BSA, (E)Hapten2b-BSA, (F) Hapten2c-BSA.

(A)

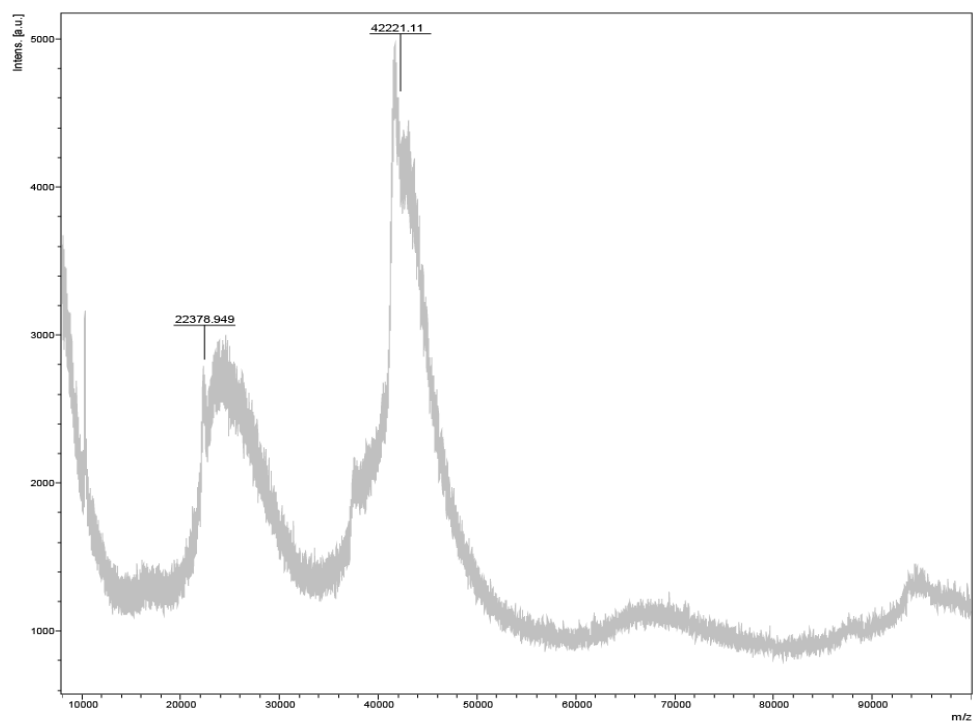

(B)

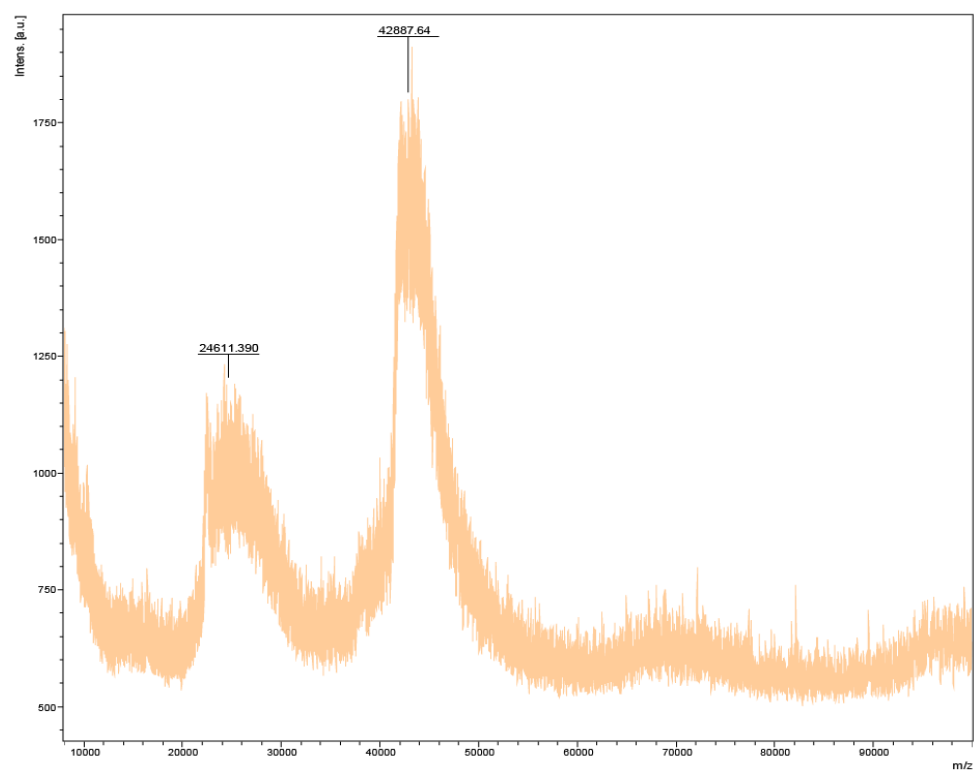

(C)

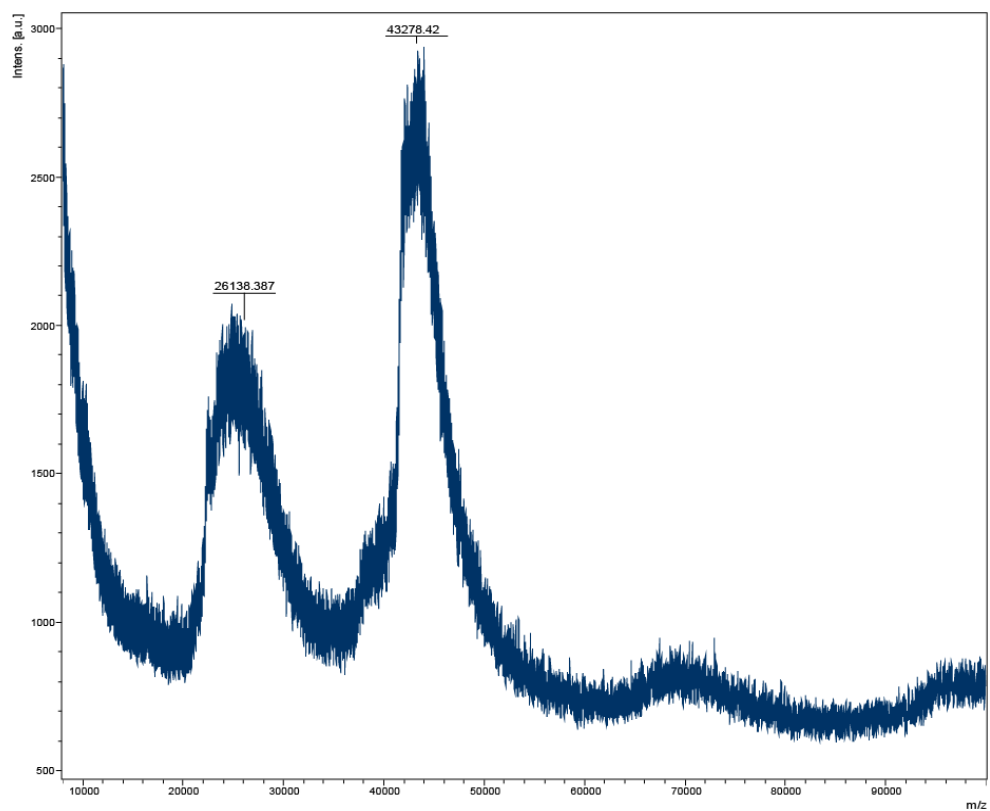

(D)

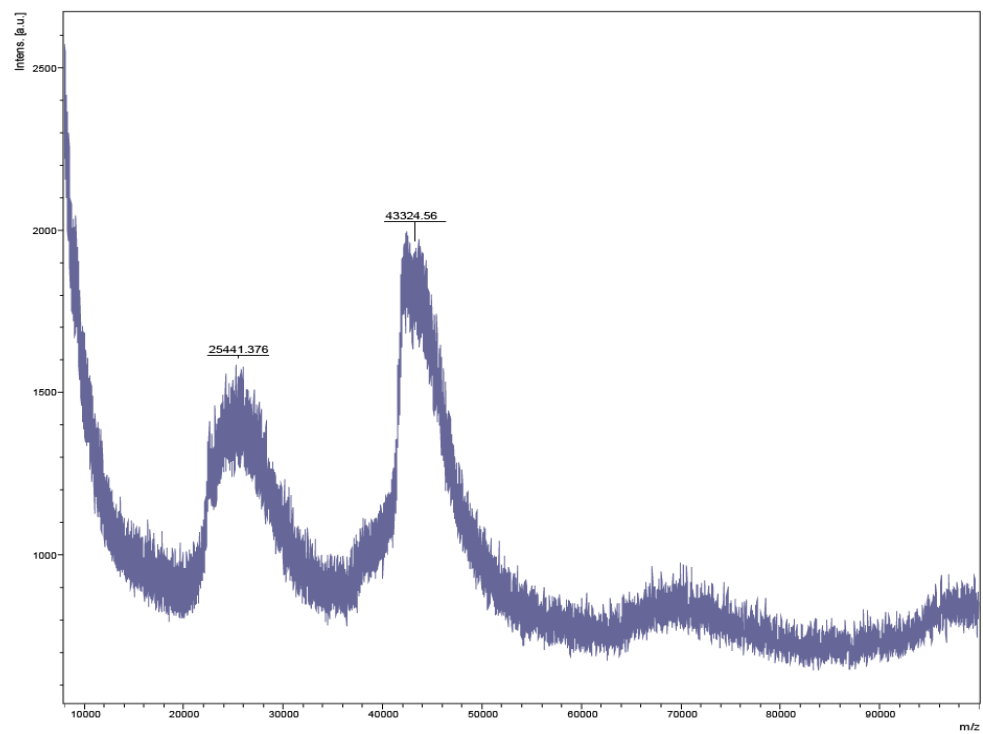

(E)

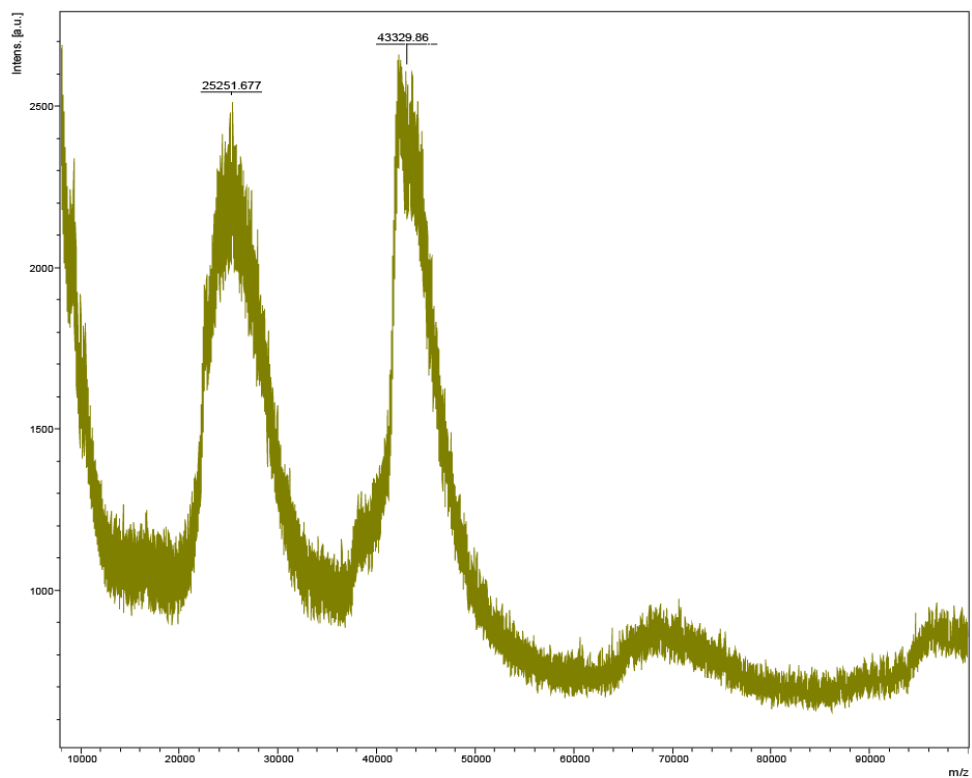

(F)

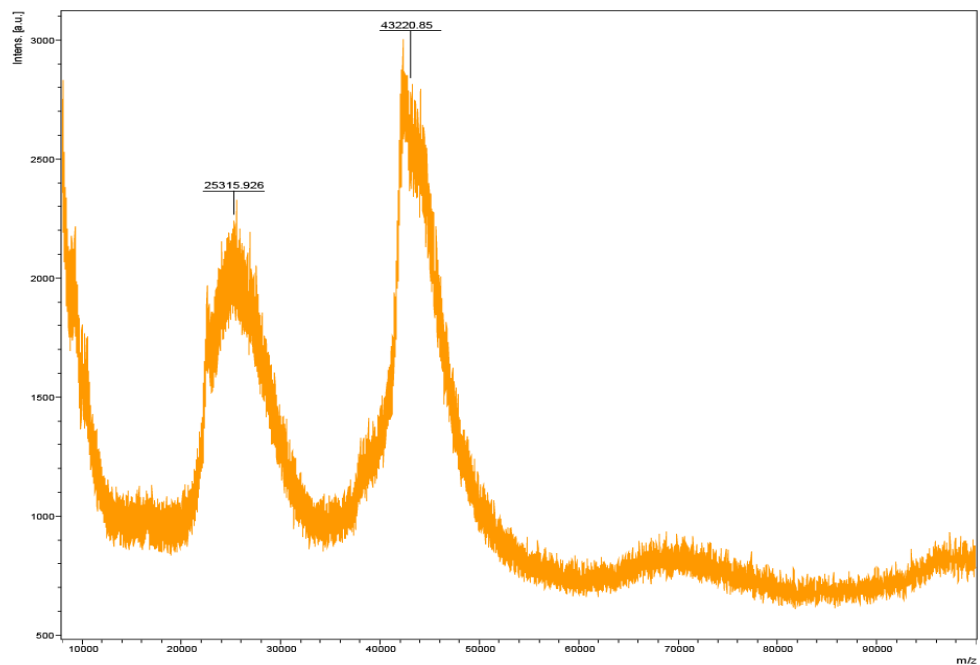

**Figure S3**. MALDI–TOF-MS of (A) OVA, (B) Hapten1a-OVA (C) Hapten1b -OVA (D) Hapten2a-OVA, (E) Hapten2b-OVA, (F) Hapten2c-OVA.

**Table S2.** The selected conjugates with similar hapten-to-protein molar ratios.

| Immunogens | ratios | Coating antigens | ratios |
|---|---|---|---|
| Hapten1a-BSA | 14.6 | Hapten2a-OVA | 5.2 |
| Hapten1b-BSA | 14.6 | Hapten2b-OVA | 4.4 |
| Hapten2a-BSA | 14.4 | Hapten2a-OVA | 5.3 |
| Hapten2b-BSA | 13.4 | Hapten2b-OVA | 5.3 |
| Hapten2c-BSA | 17.0 | Hapten2c-OVA | 5.2 |

## 6. The descriptors of haptens obtained by computational chemistry

All haptens properties excluding spacer arm and trimethoprim (TMP) were built in Gaussian 09 (Gaussian, Wallingford, CT, USA). Then, the density functional theory (DFT) calculations with the M06-2X density functional and TZVP basis set were performed to optimize the haptens [7-8]. The fundamental vibrations were calculated by the same method to confirm whether they were true minima or not. The minimum energy conformations of haptens were shown in Figure S4. The descriptors of haptens and TMP including dipole moment (μ), the energy of the highest occupied molecular orbital ($E_{HOMO}$) and of the lowest unoccupied molecular orbital ($E_{LUMO}$), molecular energy (E) were directly extracted from output file. Energy difference between the $E_{LUMO}$ and $E_{HOMO}$ ($\Delta E_{LH}$) were calculated by using Excel. Hydrophobic constant (LogP), cLogP and molecular weight (MW) were extracted by using ChemBio3D (PerkinElmer). Molecules volume ($V_m$), surface area (SA), polar surface area (PSA) and molecular polarity index (MPI) were extracted by using Multiwfn[9]. Solvent accessible surface areas (SASA) were extracted by using VMD [10].

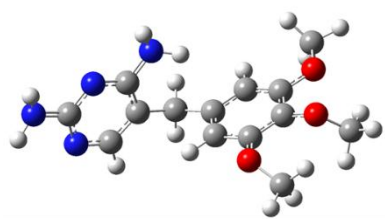 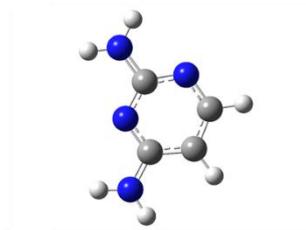 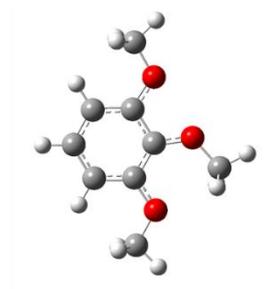

Trimethoprim　　　　　　　Hapten 1a　　　　　　　Hapten 1b

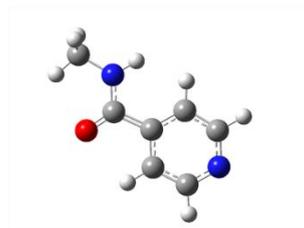 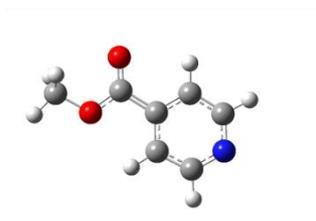 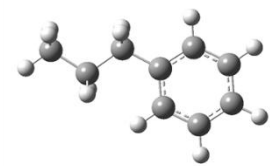

Hapten 2a　　　　　　　Hapten 2b　　　　　　　Hapten 2c

**Figure S4.** The minimum energy conformations of haptens which excluding the spacer arm. The elements are represented in the following manner: oxygen, red; nitrogen, blue; hydrogen, white; and carbon, off white.

**Table S3.** Molecular descriptors of main structure of all haptens

| Haptens [a] descriptors | Hapten1a 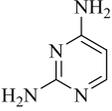 | Hapten1b 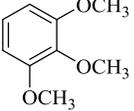 | Hapten2a 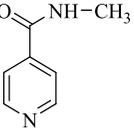 | Hapten2b 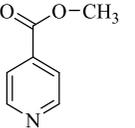 | Hapten2c 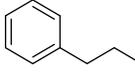 | TMP 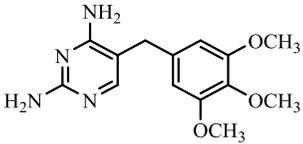 |
|---|---|---|---|---|---|---|
| μ | 2.2508 | 1.803 | 3.818 | 2.5515 | 0.4963 | 4.0933 |
| MPI | 15.22 | 11.08 | 13.66 | 11.02 | 6.90 | 12.82 |
| cLogP | -0.336 | 1.4421 | 0.108 | 0.767 | 3.699 | 0.9811 |
| LogP | -0.28 | 1.65 | -0.16 | 0.52 | 3.36 | 1.43 |
| MW | 110.12 | 168.19 | 136.15 | 137.14 | 120.2 | 290.32 |
| $E_{HOMO}$ | -0.26825 | -0.25055 | -0.32723 | -0.33154 | -0.29427 | -0.26393 |
| $E_{LUMO}$ | 0.024254 | 0.039589 | -0.0302 | -0.03798 | 0.024412 | 0.017181 |
| $\Delta E_{LH}$ | 0.292507 | 0.290136 | 0.297032 | 0.293568 | 0.31868 | 0.281107 |
| E | -375.043 | -575.756 | -456.257 | -476.128 | -350.136 | -988.928 |
| $V_m$ | 136.7412 | 214.8453 | 172.7826 | 167.8135 | 184.8898 | 353.1313 |
| SA | 145.2302 | 211.5105 | 176.7082 | 173.2873 | 185.6047 | 321.5312 |
| PSA | 95.21 | 107.46 | 94.94 | 89.86 | 41.14 | 190.87 |
| SASA | 273.0459 | 369.1366 | 313.8543 | 309.5389 | 319.3560 | 497.0549 |

[a] Here, the structures used for calculation chemistry is the subtractive haptens which excluding the spacer arm.

## 7. Immunization procedure of mice

Ten female BALB/c mice for each hapten conjugates, 6–8 weeks old, were immunized subcutaneously with each 100 μg of immunogens in 0.25 mL of 0.9% NaCl and 0.25 mL of Freund incomplete adjuvant. On the 25$^{th}$ and 50$^{th}$ days after the initial injection, mice were boosted with 100 μg of immunogen in Freund complete adjuvant. One week after injection, antisera samples obtained from the eye socket of mice were collected to detect the sensitivity and specificity based on the ELISA.

**Table S4.** The information of immunization procedures

| Immunogens | Dose/mouse | Number of immunizations | Interval time (d) | Number of mice |
| --- | --- | --- | --- | --- |
| Hapten1a-BSA | 100 μg | 3 | 25 | 10 |
| Hapten1b-BSA | 100 μg | 3 | 25 | 10 |
| Hapten2a-BSA | 100 μg | 3 | 25 | 10 |
| Hapten2b-BSA | 100 μg | 3 | 25 | 10 |
| Hapten2c-BSA | 100 μg | 3 | 25 | 10 |

## 8. The determination of titers and affinity

*8.1 The determination of titers*

The titers of antibody were determined by noncompetitive ELISA and the dilution of antibody when furnishing $OD_{max}$ between 1.5 and 2.0 as the following procedure: polystyrene 96-well microtiter plates were coated with each homogenous coating antigen (100 μL/well) incubated at 4 °C overnight. The plates were washed three times with washing buffer, and then blocked with blocking buffer (300 μL/well) at 37 °C for 1 h. Antibody (100 μL/well) were added to wells. Unbound compounds were removed by washing solution after incubation for 30 min at 37 °C. Goat anti-mouse IgG-HRP (1/5000 in PBS, 100 μL/well) were added and incubated at 37 °C for 30 min and then washed three times with washing buffer. The substrate solution (100 μL/well) was added and incubated at 37 °C for 15 min before adding 2 M $H_2SO_4$ (50 μL/well).

Absorbance values were measured at 450 nm.

*8.2 The determination of affinity*

The affinity of antibodies was determinated by using competitive indirect ELISA (ciELISA) and expressed by inhibition ratio values as followed. The microplates were firstly coated with 1μg of each homogenous coating antigen (100 μL/well) which was diluted with CB buffer and then incubated at 4 °C overnight. The coating solution in the ELISA plates was then discarded. Blocking buffer was then added to the plates (150 μL/well) which were placed in an incubator for 1 h at 37 °C. The corresponding haptens (50 μL) at 10 μg/L were added to the wells as well as 50 μL of antibody at titer dilution. After an incubation of 30 min at 37 °C, the plates were washed by PBST for two times. Goat anti-mouse IgG (HRP labeled) (1:5000, 100 μL/well) was then added and incubated for 30 min at incubator. Then 100 μL TMB substrate was added into wells and incubated for 15 min at 37 °C after washing. Then 2 mol $L^{-1}$ $H_2SO_4$ (50 μL/well) was used to stop the enzymatic reaction and the OD values of 450 nm was measured.

The inhibition ratio was calculated by using the following equation:

$$\text{Inhibition ratio} = (B_0 - B)/B_0 \qquad (\text{Eq. 2})$$

where $B_0$ and B are the responses the OD values in the absent of haptens and in the concentration of haptens at 10 mg/L.

**Table S5.** The detailed information of antibodies titers and inhibition ratios from three immunizations

| Haptens | Hapten Structures | MPI | 1st immunization | | 2nd immunization | | 3rd immunization | |
|---|---|---|---|---|---|---|---|---|
| | | | Titers [a] | Inhibition ratios [b] | Titers | Inhibition ratios | Titers | Inhibition ratios |
| Hapten1a | 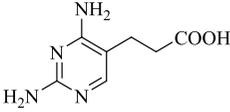 | 15.22 | 1233 | 1.09 | 3000 | 0.99 | 3360 | 0.92 |
| Hapten1b | 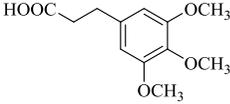 | 11.08 | 12666 | 61.02 | 28880 | 65.98 | 38400 | 77.52 |
| Hapten2a | 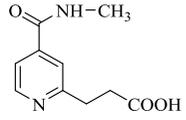 | 13.66 | 130 | 9.43 | 333 | 18.23 | 700 | 44.19 |
| Hapten2b | 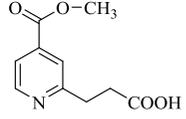 | 11.02 | 380 | 5.01 | 600 | 14.13 | 1100 | 31.09 |
| Hapten2c | 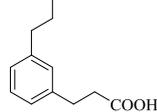 | 6.90 | 2488 | 45.42 | 5600 | 73.24 | 7111 | 90.01 |

[a] The titers of antibody were determinated by noncompetitive ELISA and the dilution of antibody when furnishing ODmax between 1.5 and 2.0. Data were average values of 10 mice and each mouse obtained in triplicate.

[b] The inhibition ratio was calculated by using the following equation: Inhibition ratio = $(B_0-B)/B_0$, where $B_0$ and B are the responses the OD values in the absent of haptens and in the concentration of haptens at 10 mg/L. Data were average values of 10 mice and each mouse obtained in triplicate.